\documentclass[10pt]{bmc_article}

\usepackage{cite} % Make references as [1-4], not [1,2,3,4]
\usepackage{url}  % Formatting web addresses  
\usepackage{ifthen}  % Conditional 
\usepackage{multicol}   %Columns
\usepackage[utf8]{inputenc} %unicode support
\usepackage{color}

\usepackage{textcomp,graphicx,epsfig}
\newboolean{publ}

\newenvironment{bmcformat}{\baselineskip15pt\sloppy\setboolean{publ}{false}}{\baselineskip15pt\sloppy}

\begin{document}
\begin{bmcformat}


\title{Large scale statistical analysis of GEO  datasets}
 
%%%%%%%%%%%%%%%%%%%%%%%%%%%%%%%%%%%%%%%%%%%%%%
%%                                          %%
%% Enter the authors here                   %%
%%                                          %%
%% Ensure \and is entered between all but   %%
%% the last two authors. This will be       %%
%% replaced by a comma in the final article %%
%%                                          %%
%% Ensure there are no trailing spaces at   %% 
%% the ends of the lines                    %%     	
%%                                          %%
%%%%%%%%%%%%%%%%%%%%%%%%%%%%%%%%%%%%%%%%%%%%%%

\author{Bernard Ycart\correspondingauthor$^{1,2,3}$%
         \email{Bernard Ycart\correspondingauthor - Bernard.Ycart@imag.fr}
       \and 
         Konstantina Charmpi$^{1,2,3}$%
         \email{Konstantina Charmpi - Konstantina.Charmpi@imag.fr}%
       \and 
         Sophie Rousseaux$^{1,4}$%
         \email{Sophie Rousseaux - Sophie.Rousseaux@ujf-grenoble.fr}%
       \and 
         Jean-Jacques Fourni\'e$^{3,5,6,7}$%
         \email{Jean-Jacques Fourni\'e - Jean-Jacques.Fournie@inserm.fr}%
      }

%%%%%%%%%%%%%%%%%%%%%%%%%%%%%%%%%%%%%%%%%%%%%%
%%                                          %%
%% Enter the authors' addresses here        %%
%%                                          %%
%%%%%%%%%%%%%%%%%%%%%%%%%%%%%%%%%%%%%%%%%%%%%%

\address{%
\iid(1) Universit\'e Grenoble Alpes, France\\
\iid(2) Laboratoire Jean Kuntzmann, CNRS UMR5224, Grenoble, France\\
\iid(3) Laboratoire d’Excellence TOUCAN, France\\
\iid(4) INSERM, UMR823, Institut Albert Bonniot, Grenoble, France\\
\iid(5) INSERM UMR1037-Cancer Research Center of Toulouse, Toulouse, France\\ 
\iid(6) Universit\'e Toulouse III Paul-Sabatier, Toulouse, France\\
\iid(7) ERL 5294 CNRS, Toulouse, France%
}%

\maketitle

%%%%%%%%%%%%%%%%%%%%%%%%%%%%%%%%%%%%%%%%%%%%%%
%%                                          %%
%% The Abstract begins here                 %%
%%                                          %%  
%% Please refer to the Instructions for     %%
%% authors on http://www.biomedcentral.com  %%
%% and include the section headings         %%
%% accordingly for your article type.       %%   
%%                                          %%
%%%%%%%%%%%%%%%%%%%%%%%%%%%%%%%%%%%%%%%%%%%%%%

\begin{abstract}
        % Do not use inserted blank lines (ie \\) until main body of text.
The problem addressed here is that of simultaneous treatment of
several gene expression datasets, possibly collected under different experimental
conditions and/or platforms. Using robust statistics, a
large scale statistical analysis has been conducted over $20$ datasets 
downloaded from the Gene Expression Omnibus repository. The differences
between datasets are compared to the variability inside a given
dataset. Evidence that meaningful biological 
information can be extracted by
merging different sources is provided.
\end{abstract}

\ifthenelse{\boolean{publ}}{\begin{multicols}{2}}{}

%%%%%%%%%%%%%%%%%%%%%%%%%%%%%%%%%%%%%%%%%%%%%%
%%                                          %%
%% The Main Body begins here                %%
%%                                          %%
%% Please refer to the instructions for     %%
%% authors on:                              %%
%% http://www.biomedcentral.com/info/authors%%
%% and include the section headings         %%
%% accordingly for your article type.       %% 
%%                                          %%
%% See the Results and Discussion section   %%
%% for details on how to create sub-sections%%
%%                                          %%
%% use \cite{...} to cite references        %%
%%  \cite{koon} and                         %%
%%  \cite{oreg,khar,zvai,xjon,schn,pond}    %%
%%  \nocite{smith,marg,hunn,advi,koha,mouse}%%
%%                                          %%
%%%%%%%%%%%%%%%%%%%%%%%%%%%%%%%%%%%%%%%%%%%%%%

%%%%%%%%%%%%%%%%
%% Background %%
%%
\section*{Background}
Many genomewide expression datasets have been published during the
past ten years. 
Repositories, such as the Gene Expression Omnibus (GEO)
database \cite{Edgar02}, have made
available an impressive wealth of data. Using
them as a whole, instead of restricting statistical studies 
to one particular dataset, is tantalizing. Two recently published 
R/Bioconductor packages \cite{Heider_14,Taminau_inSilicoMerging}
provide various
tools for merging datasets coming from different studies.
However, a serious doubt has been cast by Haibe-Kains et al. 
\cite{Haibe-Kains13}, after comparing two large scale pharmacogenomic
studies: whereas both studies had a good overall correlation,  
important discordances could be observed. Thus, the following crucial
question remains to be answered: is it statistically legitimate to
merge datasets coming from different studies? An attempt at answering
this question is reported here.
 
Merging different datasets, requires prior checking
that the information they contain is compatible, and hence that detected
differences between gene expressions under different conditions 
are not artifacts, due to experimental or data
processing methods.
An obvious obstacle to simultaneous treatment is that
expression data collected under different experimental conditions
and/or platforms usually have incompatible distributions, which differ
sometimes by several orders of magnitude \cite{Kuo02,Mah04}. A solution is
provided by robust (or distribution-free) statistics 
\cite{GibbonsChakraborti03,Heritieretal09}. Robust methods 
amount to replacing actual
values by ranks, or equivalently by empirical distribution functions
or van der Waerden's normal scores
\cite[p.~309]{GibbonsChakraborti03}. This
idea has already been applied to expression data 
in several papers, including
\cite{Tsodikov02,Warnat05,Breitling04}. However,
to the best of our knowledge, a large scale analysis assessing the
reproducibility of information from one dataset to another, is still
missing. We have conducted such an analysis over
$20$ GEO datasets, totalling $17\,745$ 
genomewide expression samples.

For the data treatments  presented here, the
statistical language R \cite{R_software}  has been used. Our set of
functions, together with a manual, has been made
available online as supplementary information.
Throughout the article, we consider \emph{data matrices} (also called
assay data in  \cite{Gentleman04}) as containing
expression data relative to a set of genes.  
Each row corresponds to a different gene symbol, or \emph{feature}, 
each column to a
different data vector or \emph{sample} (see Table 2 in \cite{Edgar02}). 
Such a matrix is
deduced from raw datasets, available on the GEO
repository, though standard treatments: annotation and reduction
\cite{Gentleman_genefilter}.
Several R packages \cite{Davis07,Taminau11} that
perform these operations and output data matrices such as
considered here, are available. We have encoded our own functions. 
We have chosen a data structure in which each data
matrix is paired with its \emph{information matrix}. The columns of the
information matrix are labelled by the same numbers as the paired data
matrix. Its rows contain the different information fields of the
data. Our focus here is on overexpression or underexpression of genes, in
different tissues or cancer types.

Our objective was twofold. On the one hand, we wanted to check whether
the information on genes, contained in different data matrices, was
compatible, and to which extent. This was done on a set of $20$ different
matrices. Various statistical treatments were performed. 
The first one consisted in computing
correlations between median columns of the matrices. Vectors of
pairwise correlations between rows were also compared. Then
multivariate analysis over assays of gene symbols was applied:
Wilcoxon and Kruskal-Wallis tests, factor and principal component
analysis (PCA). The results
were compared to those obtained by sorting a single
matrix according to different keywords. All
comparisons showed not perfect, but highly significant
correlations. However, it was also found that in all cases, a sizeable proportion
of symbols were good discriminators of the different matrices. But
this was also found to hold between two submatrices inside a given
dataset. Therefore, it cannot be regarded as an obstacle to merging
different datasets. 
On the other hand, we wanted to know whether biological information could
be consistently retrieved from matrices collated from different
sources. Two merged sets of matrices were made. The first one came
from general cancer cell datasets, from which samples
of breast and lung tumors were extracted. The second one was made of
blood RNA datasets, coming from healthy individuals, or from
leukemias.  In both cases, evidence that already known 
biological information could be extracted from
merged matrices was found. 
\section*{Methods}
The datasets available from the GEO repository \cite{Edgar02},
collate sets of expression vectors, or samples. 
Several R/Bioconductor packages can
be used to download and format the data \cite{Davis07,Taminau11}. We have chosen
to encode in R our own functions. Our R script has been made available
online, together with a user manual. Our formatting choices are
described below. 

In a GEO dataset two types of information are available for
each sample. The first type consists of numeric values corresponding to a set
of probes. The second type are character-type informations on the
experimental setting. We have chosen to separate the two types into
data matrices and information matrices. In the data matrix, probes are
associated to gene symbols with the use of different Bioconductor annotation 
packages according to the platform 
\cite{Carlson_hgu133plus2.db,Carlson_hgu133a.db,
Dunning_illuminaHumanv3.db,Dunning_illuminaHumanv4.db}. After
annotation, some symbols are 
duplicated. Several methods can be used to eliminate duplicates. We
have chosen to keep the row with the largest interquartile range, as in
\cite{Gentleman_genefilter}, because we believe that this is the most
statistically coherent 
choice. After annotation and reduction, the data matrix, with
gene symbols as row names, and series numbers as column names, is saved
as a single R object for future use. The information matrix has the
same column names as the corresponding data matrix. Its
rows correspond to the different fields.

Our merging function reduces
 data matrices to common row names. For information matrices,
different sets of data usually have different
information fields. This was taken into account when merging two information
matrices, by indexing the rows of the merged matrix by the
union of row names in the initial information matrices.    

Two R/Bioconductor packages have recently been issued for merging GEO datasets 
\cite{Heider_14,Taminau_inSilicoMerging}. 
In \cite{Heider_14}, quantile discretization,
normal discretization normalization, 
gene quantile normalization, median rank scores, 
quantile normalization (QN) are proposed. In
\cite{Heider_14,Taminau_inSilicoMerging}, 
the Batch Mean-Centering method, Distance-Weighted Discrimination,
Z-score standardization, and the Cross-Platform Normalization method
are proposed. An Empirical Bayes (EB) method is available in both
packages. For the results reported here, only classical methods were
used, and we consider them as sufficient to establish our main points,
our focus being on overexpression or underexpression of genes, in
different tissues or cancer types.

As in \cite{Tsodikov02,Warnat05,Breitling04}, 
we have made the choice to use robust statistics 
\cite{GibbonsChakraborti03,Heritieretal09}. This implies changing
the columns of a data matrix into distribution free values. The usually
proposed transformation replaces the $i$-th value $x_i$ 
by its rank $R_i$ if $x_i$ is the $R_i$-th smallest value in the
column. However, ranks range between $1$ and the number of
rows. The problem is that different matrices may have different
numbers of rows (gene symbols). In order to get a unique range of
values for all matrices, it seems preferable to use a scale free
score. The simplest such score is the Empirical Cumulative 
Distribution Function (ECDF):
its value at $x_i$ is $R_i/n$, if $n$ is the number of rows. Graphical
displays look more familiar if another score is
used: the van der Waerden's normal score
\cite[p.~309]{GibbonsChakraborti03}. It consist of replacing
$x_i$ by $\phi(R_i/(n+1))$, where $\phi$ is the quantile function of the
standard normal distribution. With the ECDF, the distribution of each
column becomes uniform on the 
interval $[0,1]$, whereas with the normal score it becomes standard
normal. Results reported above have been obtained with the normal
score, but they are not essentially different if the ECDF is
used instead.

In statistical inference, 
the choice of robust statistics must be made coherent. This is
the reason why we have replaced the usual normal-sample techniques by
their robust equivalent, and used medians instead of means,
Spearman's correlation instead of
Pearson's \cite[p.~422-431]{GibbonsChakraborti03}, Wilcoxon (or
Mann-Whitney) location test instead of 
Student's t-test \cite[p.~268-278]{GibbonsChakraborti03}, Kruskal-Wallis
test instead of one-way analysis of variance
\cite[p.~363-372]{GibbonsChakraborti03}. When comparing several
matrices to detect location diffferences, 
the Kruskal-Wallis test was run over all common rows. When
differentiating overexpression from underexpression, a one-sided
Wilcoxon test was run. The same test being used for a large number of
features, a False Detection Rate (FDR) 
correction of p-values by the Benjamini-Yekutieli method
\cite{Benjamini_Yekutieli01} was systematically applied. Features were
ranked from most to least significant, either by sorting p-values in
increasing order, or by sorting the values of the test statistic
instead. We considered as significant, any feature with a
(FDR-corrected) p-value smaller than 5\%.
Once a set of (significant) features had been selected, 
the corresponding rows were concatenated into single vectors. These
vectors were taken as variables, and the samples as individuals, 
for a PCA. Figures 3 to 5 were obtained by
projecting the samples as points onto the first principal plane,
and differentiating their initial data matrices by colors. Precise
R commands can be found in the user manual made available online. 
\section*{Results}
The $20$ datasets that were downloaded from the GEO repository are
detailed in Table 1. They were selected on a criterion of size (number of
samples: $500$ or more). The $20$ matrices together amount
to $17\,745$ samples. To each study, a three-letter acronym was
attached; these acronyms
will be used in what follows. 

\begin{table}[!ht]
\begin{center}
\begin{tabular}{|c||ccccc|}
        \hline 
acronym&reference &series number&platform number&symbols (rows)&samples (columns) \\\hline
EPO&\cite{expO}&2109&570&20\,184&2158\\
PMM&\cite{Chen10}&2658&570&20\,184&559\\
AML&\cite{Jonge10}&6891&570&20\,184&537\\
HBI&\cite{Roth07}&7307&570&20\,184&677\\
MIL&\cite{Haferlach10}&13159&570&20\,184&2\,096\\
MDS&\cite{Mills09}&15061&570&20\,184&870\\
PLE&\cite{Fehrmann11}&20142&6947&19\,626&1\,240\\
MMD&\cite{Popovici10}&24080&570&20\,184&559\\
DLB&\cite{Frei13}&31312&570&20\,184&498\\
PRS&\cite{Westra13}&33828&10558&20\,768&881\\
CCL&\cite{Barretina12}&36133&15308&18\,722&917\\
BEC&\cite{Hernandez12}&36192&6947&19\,628&911\\
WBS&\cite{Mayerle13}&36382&6947&19\,628&991\\
GSC&\cite{Xiao11}&36809&570&18\,260&812\\
MBI&\cite{Seok13}&37069&570&18\,260&590\\
CCC&\cite{Marisa13}&39582&570&20\,184&566\\
PVA&\cite{Wood11}&48152&6947&19\,628&705\\
HPS&\cite{Esko13}&48348&6947&19\,628&734\\
XMD&\cite{Hollingshead13}&48433&570&20\,184&823\\
HAV&\cite{Obermoser13}&48762&6947&19\,628&621\\\hline
      \end{tabular}
\caption{Twenty GEO series have been chosen, coming from four different
platforms. To each of them a three letters acronym was associated. The
table gives the acronym, a recent reference, the GEO series number, 
the platform number. For the data matrix (or assayData), 
the number of symbols after annotation and reduction, and
the number of columns (samples) are given. All
$20$ data matrices had $15\,562$ gene symbols in common.}
\end{center}
\end{table}

In the results reported
here, each data matrix has been 
transformed by replacing its column values, by the corresponding
van der Waerden normal scores
\cite[p.~309]{GibbonsChakraborti03}. Similar results were obtained when
replacing column values by their empirical distribution function
(see methods section).

The first treatment that was applied
consisted in computing, for each dataset, the median of all rows,
reduced to the $15\,562$ common gene symbols. 
 This
gave $20$ vectors of length $15\,562$, the correlation matrix of which
is given in Table 2. A positive (negative) 
correlation between vectors of size $15\,562$
is significant at threshold 5\% if it is larger than $0.013$ (smaller
than $-0.013$); thus all correlations of Table 2 can be regarded as
significant.

\begin{table}[!ht]
\begin{center}
{\footnotesize
\begin{tabular}{|r||rrrrrrrrrr|}
       \hline
     &EPO  &PMM  &AML  &HBI  &MIL  &MDS  &PLE  &MMD  &DLB  &PRS  \\ \hline
EPO &$1.00 $&$0.80 $&$0.71 $&$0.92 $&$0.63 $&$0.82 $&$0.55 $&$0.48 $&$0.59 $&$0.18$ \\
PMM &$0.80 $&$1.00 $&$0.75 $&$0.79 $&$0.63 $&$0.81 $&$0.56 $&$0.63 $&$0.55 $&$0.19$ \\
AML &$0.71 $&$0.75 $&$1.00 $&$0.68 $&$0.76 $&$0.85 $&$0.61 $&$0.58 $&$0.59 $&$0.18$ \\
HBI &$0.92 $&$0.79 $&$0.68 $&$1.00 $&$0.58 $&$0.76 $&$0.53 $&$0.46 $&$0.53 $&$0.18$ \\
MIL &$0.63 $&$0.63 $&$0.76 $&$0.58 $&$1.00 $&$0.78 $&$0.58 $&$0.67 $&$0.61 $&$0.14$ \\
MDS &$0.82 $&$0.81 $&$0.85 $&$0.76 $&$0.78 $&$1.00 $&$0.69 $&$0.52 $&$0.60 $&$0.18$ \\
PLE &$0.55 $&$0.56 $&$0.61 $&$0.53 $&$0.58 $&$0.69 $&$1.00 $&$0.40 $&$0.45 $&$0.22$ \\
MMD &$0.48 $&$0.63 $&$0.58 $&$0.46 $&$0.67 $&$0.52 $&$0.40 $&$1.00 $&$0.50 $&$0.11$ \\
DLB &$0.59 $&$0.55 $&$0.59 $&$0.53 $&$0.61 $&$0.60 $&$0.45 $&$0.50 $&$1.00 $&$0.12$ \\
PRS &$0.18 $&$0.19 $&$0.18 $&$0.18 $&$0.14 $&$0.18 $&$0.22 $&$0.11 $&$0.12 $&$1.00$ \\
CCL &$0.81 $&$0.74 $&$0.69 $&$0.75 $&$0.65 $&$0.77 $&$0.55 $&$0.54 $&$0.56 $&$0.20$ \\
BEC &$0.56 $&$0.49 $&$0.43 $&$0.65 $&$0.38 $&$0.48 $&$0.63 $&$0.32 $&$0.35 $&$0.21$ \\
WBS &$0.57 $&$0.58 $&$0.62 $&$0.54 $&$0.58 $&$0.70 $&$0.94 $&$0.40 $&$0.46 $&$0.23$ \\
GSC &$0.59 $&$0.61 $&$0.69 $&$0.57 $&$0.63 $&$0.74 $&$0.66 $&$0.46 $&$0.49 $&$0.15$ \\
MBI &$0.60 $&$0.63 $&$0.70 $&$0.58 $&$0.65 $&$0.75 $&$0.64 $&$0.48 $&$0.50 $&$0.15$ \\
CCC &$0.77 $&$0.62 $&$0.64 $&$0.67 $&$0.59 $&$0.67 $&$0.50 $&$0.49 $&$0.53 $&$0.15$ \\
PVA &$-0.09 $&$-0.10 $&$-0.13 $&$-0.09 $&$-0.14 $&$-0.16 $&$-0.30 $&$-0.09 $&$-0.09 $&$0.16$\\
HPS &$0.62 $&$0.62 $&$0.66 $&$0.58 $&$0.62 $&$0.74 $&$0.93 $&$0.43 $&$0.49 $&$0.24$ \\
XMD &$0.90 $&$0.81 $&$0.72 $&$0.84 $&$0.65 $&$0.83 $&$0.56 $&$0.51 $&$0.56 $&$0.19$ \\
HAV &$0.60 $&$0.60 $&$0.64 $&$0.56 $&$0.61 $&$0.73 $&$0.89 $&$0.42$&$0.48 $&$0.21$ \\
\hline\hline
    &CCL &BEC &WBS &GSC &MBI &CCC &PVA &HPS &XMD &HAV \\\hline
EPO &$0.81 $&$0.56 $&$0.57 $&$0.59 $&$0.60 $&$0.77 $&$-0.09 $&$0.62 $&$0.90 $&$0.60$ \\
PMM &$0.74 $&$0.49 $&$0.58 $&$0.61 $&$0.63 $&$0.62 $&$-0.10 $&$0.62 $&$0.81 $&$0.60$ \\
AML &$0.69 $&$0.43 $&$0.62 $&$0.69 $&$0.70 $&$0.64 $&$-0.13 $&$0.66 $&$0.72 $&$0.64$ \\
HBI &$0.75 $&$0.65 $&$0.54 $&$0.57 $&$0.58 $&$0.67 $&$-0.09 $&$0.58 $&$0.84 $&$0.56$ \\
MIL &$0.65 $&$0.38 $&$0.58 $&$0.63 $&$0.65 $&$0.59 $&$-0.14 $&$0.62 $&$0.65 $&$0.61$ \\
MDS &$0.77 $&$0.48 $&$0.70 $&$0.74 $&$0.75 $&$0.67 $&$-0.16 $&$0.74 $&$0.83 $&$0.73$ \\
PLE &$0.55 $&$0.63 $&$0.94 $&$0.66 $&$0.64 $&$0.50 $&$-0.30 $&$0.93 $&$0.56 $&$0.89$ \\
MMD &$0.54 $&$0.32 $&$0.40 $&$0.46 $&$0.48 $&$0.49 $&$-0.09 $&$0.43 $&$0.51 $&$0.42$ \\
DLB &$0.56 $&$0.35 $&$0.46 $&$0.49 $&$0.50 $&$0.53 $&$-0.09  $&$0.49 $&$0.56 $&$0.48$ \\
PRS &$0.20 $&$0.21 $&$0.23 $&$0.15 $&$0.15 $&$0.15 $&$0.16 $&$0.24 $&$0.19 $&$0.21$ \\
CCL &$1.00 $&$0.56 $&$0.56 $&$0.56 $&$0.59 $&$0.74 $&$-0.08 $&$0.61 $&$0.92 $&$0.60$ \\
BEC &$0.56 $&$1.00 $&$0.64 $&$0.37 $&$0.37 $&$0.45 $&$-0.13 $&$0.65 $&$0.57 $&$0.59$ \\
WBS &$0.56 $&$0.64 $&$1.00 $&$0.64 $&$0.63 $&$0.49 $&$-0.29 $&$0.95 $&$0.57 $&$0.88$ \\
GSC &$0.56 $&$0.37 $&$0.64 $&$1.00 $&$0.94 $&$0.61 $&$-0.19 $&$0.66 $&$0.57 $&$0.64$ \\
MBI &$0.59 $&$0.37 $&$0.63 $&$0.94 $&$1.00 $&$0.62 $&$-0.17 $&$0.66 $&$0.59 $&$0.64$ \\
CCC &$0.74 $&$0.45 $&$0.49 $&$0.61 $&$0.62 $&$1.00 $&$-0.09 $&$0.53 $&$0.75 $&$0.51$ \\
PVA &$-0.08 $&$-0.13 $&$-0.29 $&$-0.19 $&$-0.17 $&$-0.09 $&$1.00 $&$-0.27 $&$-0.08 $&$-0.26$ \\
HPS &$0.61 $&$0.65 $&$0.95 $&$0.66 $&$0.66 $&$0.53 $&$-0.27 $&$1.00 $&$0.62 $&$0.90$ \\
XMD &$0.92 $&$0.57 $&$0.57 $&$0.57 $&$0.59 $&$0.75 $&$-0.08 $&$0.62 $&$1.00 $&$0.60$ \\
HAV &$0.60 $&$0.59 $&$0.88 $&$0.64 $&$0.64 $&$0.51 $&$-0.26 $&$0.90 $&$0.60 $&$1.00$ \\\hline
\end{tabular}
}
\caption{For each of the $20$ data matrices of Table 1, the
median column value of each gene symbol was computed. This gave $20$
vectors with length  $15\,562$ (number of common symbols). The 
table gives pairwise correlations between the $20$ vectors.}
\end{center}
\end{table}

Figure 1 shows a factor analysis of the 20
variables. Fifteen of them can be clustered into four groups.
%-------------------------------------------------------------------
\begin{figure}[!ht]
\centerline{
\includegraphics[width=10cm]{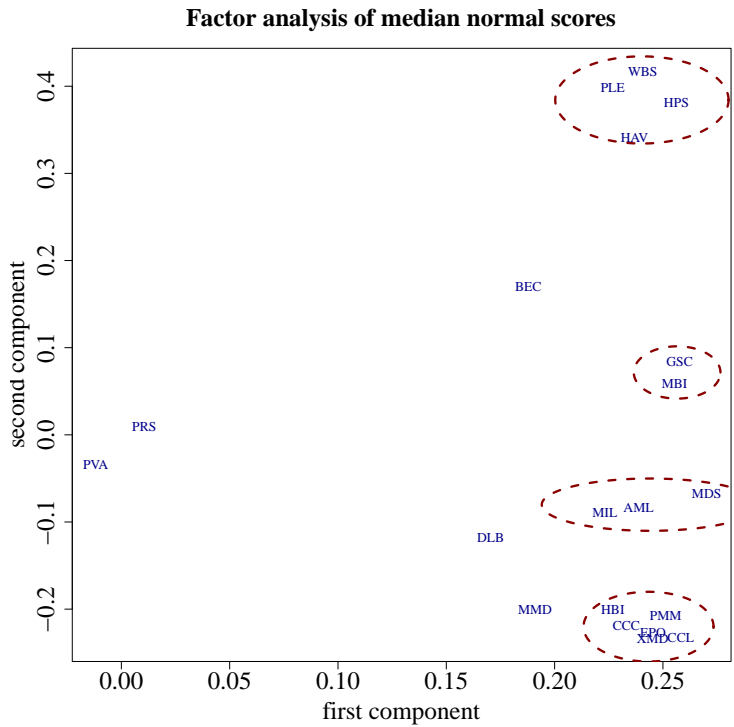}
} 
\caption{Factor analysis of median columns for 20 datasets. The 20
  variables are projected onto the first principal plane of the
  PCA. Four clusters are identified.}
\label{fig:1}
\end{figure}
%-------------------------------------------------------------------
\begin{itemize}
\item PMM, EPO, XMD, HBI, CCL, CCC. Among these six datasets, four are
  generalist studies involving different tissues and conditions (EPO,
  HBI, XMD, CCL); CCC concerns colon cancer, and PMM multiple
  myelomas. Observe that CCL, which was obtained under a platform
  different from the five others, has excellent correlations with
  them (between 0.74 and 0.92).
\item WBS, PLE, HPS, HAV. All four correspond to blood RNA samples
  from healthy patients.
\item MIL, AML, MDS. All three correspond to leukemias.
\item GSC, MBI. These two matrices correspond to similar tissues
  (blood samples), and similar conditions (critical injuries and burn
  injuries). Moreover, they were produced on the same platform, by the
  same organization. Their excellent correlation (0.94) is not a
  surprise.
\end{itemize}
Three datasets, BEC, DLB, MMD have relatively good correlations with
those of the above four groups (around 0.5), but no particular links
with those groups, nor between themselves. The
relative surprise comes from the weak correlations of PRS, and the
negative correlations of PVA. Both come from blood RNA samples, and they
could have been expected to be close to the WBS, PLE, HPS, HAV group.
That PRS and PVA are far from any other matrix, 
can be explained by their inner heterogeneity. It is illustrated for PVA
on Figure 2, where the values over 
features ALPP and CA4 are represented: samples separate into 4
clusters, according to over- or underexpression of the two
genes. As an example, if PVA is split into samples for which the value of ALPP is
positive (overexpression), or negative (underexpression), and the
row medians are calculated over the two submatrices as before, a
correlation of $-0.69$ is found: thus one half of PVA has a strong
negative correlation with the other half. Similar results are obtained for
many other features. We considered that the heterogeneity of PVA and PRS did
not qualify them for merging.
%
%-------------------------------------------------------------------
\begin{figure}[!ht]
\centerline{
\includegraphics[width=10cm]{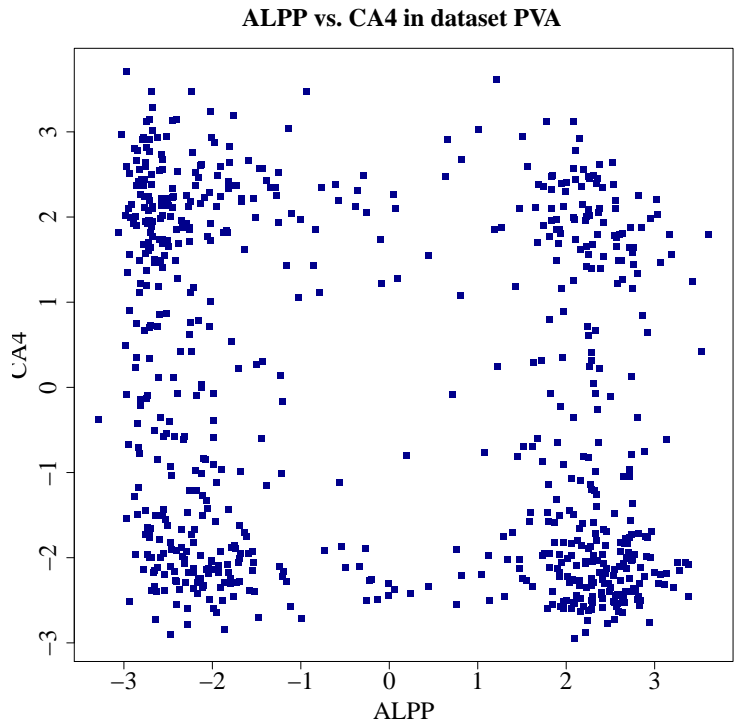}
} 
\caption{Values of PVA on CA4 versus ALPP. Samples separate into 4
  clusters, according to over- or underexpression of the two genes.}
\label{fig:2}
\end{figure}
%-------------------------------------------------------------------

For each matrix, we also computed all possible pairwise row correlations:
$20$ vectors of more than $121$ millions of pair-correlations were obtained: this
is the technique used to evaluate genes for cross-platform consistency 
of expression patterns in \cite{Parmigiani04}. As expected, the
correlation matrix had smaller values than that of Table 2. For
instance, the correlation of CCL with XMD was $0.53$ instead of
$0.92$, but still highly significant because of the large number of
values.

Correlations between column medians or pair-correlations,
is too crude a criterion to judge
the homogeneity of two datasets. As an example, GSC and MBI
have an excellent median correlation of 0.94, and several
good reasons to be similar. Yet, when each feature is tested for
significant differences by the Kruskal-Wallis test, 
$14\,800$ significant features out of $18\,260$ are detected (see methods
section for details). The same occurred for any pair of datasets: the
distributions of rows had significantly different location parameters, for a
majority of features. This means that, for a majority of genes, the
ranks of their expressions in the first dataset are significantly smaller
or larger than in the second.

Since discrepancies appear to be observed between any two datasets,
it must be decided whether they are due to actual biological
information, or to a statistical artifact, induced by 
the experimental setting or the
platform. For this, we focused on the dataset MIL
(GSE13159  \cite{Haferlach10}), that has 2\,096 samples. The samples 
were sorted into six submatrices, according to six keywords: Healthy (74 samples),
ALL (acute lymphoblastic leukemia, 750 samples),  
AML (acute myeloid leukemia, 542 samples),  
CLL (chronic lymphocytic leukemia, 448 samples),  
CML (chronic myelogenous leukemia, 750 samples),  
MDS (myelodysplastic syndrome, 202 samples). Then the same treatments as
before were applied. Firstly the six median columns were computed, and
their correlation matrix was obtained (Table 3). 
\begin{table}[!ht]
\begin{center}
\begin{tabular}{|l||cccccc|}
\hline
&        Healthy&  ALL&  AML&  CLL&  CML&  MDS\\\hline
Healthy&    1.00& 0.92& 0.96& 0.86& 0.98& 0.99\\
ALL&        0.92& 1.00& 0.95& 0.91& 0.90& 0.91\\
AML&        0.96& 0.95& 1.00& 0.89& 0.96& 0.97\\
CLL&        0.86& 0.91& 0.89& 1.00& 0.84& 0.85\\
CML&        0.98& 0.90& 0.96& 0.84& 1.00& 0.98\\
MDS&        0.99& 0.91& 0.97& 0.85& 0.98& 1.00\\\hline
\end{tabular}
\caption{The data matrix MIL was partitioned according to the 6 keywords
Healthy, ALL, AML, CLL, CML, MDS. 
For each of the six submatrices,  the
median column of each feature was computed. This gave 6 
vectors with length  $20\,184$ (number of symbols in MIL). The
table gives the correlations of the 6 vectors. 
}
\end{center}
\end{table}

The values are
 between $0.85$ and $0.99$, which is in the range of the best
correlations of Table 2. As a control, we made a partition of
the same matrix into 6 random subsets, with the same numbers of
samples as above, and computed the correlation matrix in the same way. On the
control random partition, all correlations were above $0.997$. This proves
that the partition into keywords does contain meaningful
differences. Indeed, these differences were detected by the
Kruskal-Wallis test: out of the $20\,184$ features, $18\,301$ were
found significant. Twenty-two features had
Kruskal-Wallis p-value below $10^{-300}$:
SOX4, SYNGR2, ERLIN1, FAH, C7orf23, PSMA6, RTN3, UHRF1, ADAM28,
BLK, FUCA2, CD79A, ADA, MYL6B, HEBP1, LEF1-AS1, LEF1, AFF3,    
COL9A2, MICALL2, MPO, PPM1K.
 A PCA of the corresponding
rows of MIL was run, and the samples projected as points onto the
first principal plane, differentiating submatrices by colors (Figure
3). The two submatrices ALL (blue points) and CLL (brown points) are
clearly separated from the rest.
%-------------------------------------------------------------------
\begin{figure}[!ht]
\centerline{
\includegraphics[width=10cm]{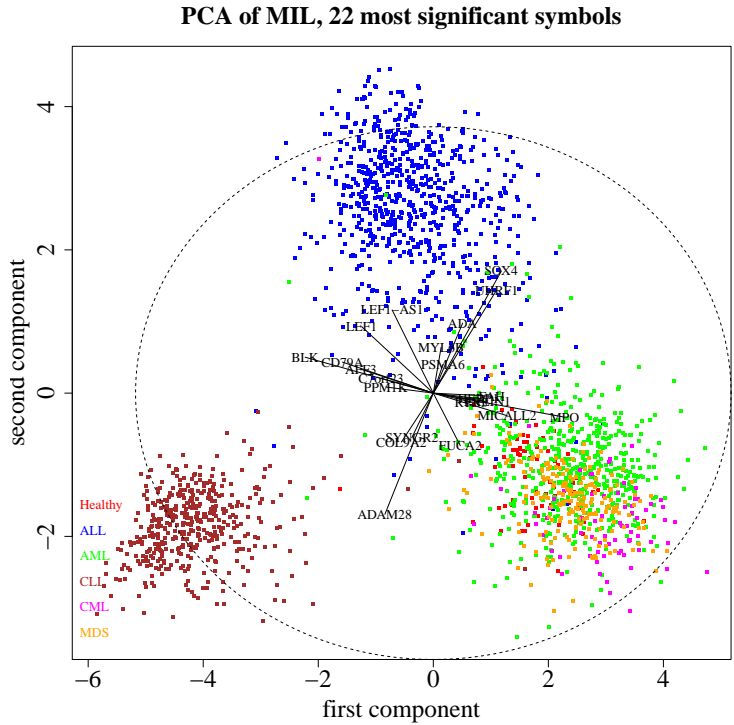}
} 
\caption{Dataset MIL, partitioned into 6 submatrices according to
  keywords Healthy, ALL, AML, CLL, CML, MDS.
PCA of the 22 symbols with
  Kruskal-Wallis p-value under $10^{-300}$: SOX4, SYNGR2, ERLIN1, FAH,
C7orf23, PSMA6, RTN3, UHRF1, ADAM28, 
BLK, FUCA2, CD79A, ADA, MYL6B, HEBP1, LEF1-AS1, LEF1, AFF3,    
COL9A2, MICALL2, MPO, PPM1K.
Samples are represented by points, with six different
colors.
}
\label{fig:3}
\end{figure}
%-------------------------------------------------------------------
 
Differences inside a given dataset can be induced by
several factors. Two factors may not induce differences of the same order of
magnitude. However, 
there is no statistical reason why a dataset like 
MIL should not be used as a whole,
and many ways to verify that the observed differences correspond to
actual biological information. Here is an example. Stirewalt et al. 
\cite{Stirewaltetal08}
%http://www.ncbi.nlm.nih.gov/pubmed/17910043
list a group of 7 genes displaying increased expression in acute
myeloid leukemia
samples: BIK, CCNA1, FUT4, IL3RA, HOMER3, JAG1, WT1. 
When a one-sided Wilcoxon test is applied to the submatrix AML versus the
rest of MIL, those 7 genes are among the most significant: their
p-values range between $6.7\times 10^{-130}$ and $4.3\times 10^{-42}$.
The most significant, HOMER3, ranks 54-th among the $20\,184$ features
of MIL. 

If observed differences between two datasets (like GSC and MBI) 
are of the same order of magnitude as differences inside a given
dataset, such as caused by a significant factor (see figure 3), 
it can be admitted as statistically legitimate to
merge the two datasets. That meaningful information can be obtained
from the merging, remains to be proved. In the following experiments,
matrices to be merged were selected in the clusters detected by factor
analysis (Figure 1).  

Our first experiment consisted in extracting samples corresponding to
breast and lung tumors, from the three matrices CCL, EPO, and XMD.
CCL has 56 samples of breast tumors, and 166 of lung tumors, EPO has
367 and 143, XMD has 32 and 152. Two matrices ``Breast'' and ``Lung''
were made by merging the six submatrices three by three, according to
tissues. They had $18\,466$ features in common, by $455$ samples for
Breast, and $461$ for Lung.

The Kruskal-Wallis test was run on
the six separated submatrices, then on the two matrices Breast and
Lung. The ten most significant symbols
were extracted, and a PCA was run as before. 
The results are displayed on Figure 4. Significant
symbols when the 6 matrices are separated (left panel) are different from
significant symbols separating Breast and Lung (right panel). On the
left panel, it is clear that the information on the dataset (CCL, EPO,
or XMD) dominates the separation Breast vs. Lung: samples coming from CCL
are on the left, from EPO on the right, from CCL
in the middle. 
But on the right panel, the two types of tumors are also
clearly separated. Separators include GATA3 on the right side
(Breast), IGF2BP3 on the left side (Lung). 
Two articles, among others, show the importance of GATA3 for breast
cancer \cite{Zakaria10,Ma13}. In \cite{Beljan12}, 
the link of IGF2BP3 to lung cancer is explicitly stated.
%-------------------------------------------------------------------
\begin{figure}[!ht]
\centerline{
\begin{tabular}{cc}
\includegraphics[width=8cm]{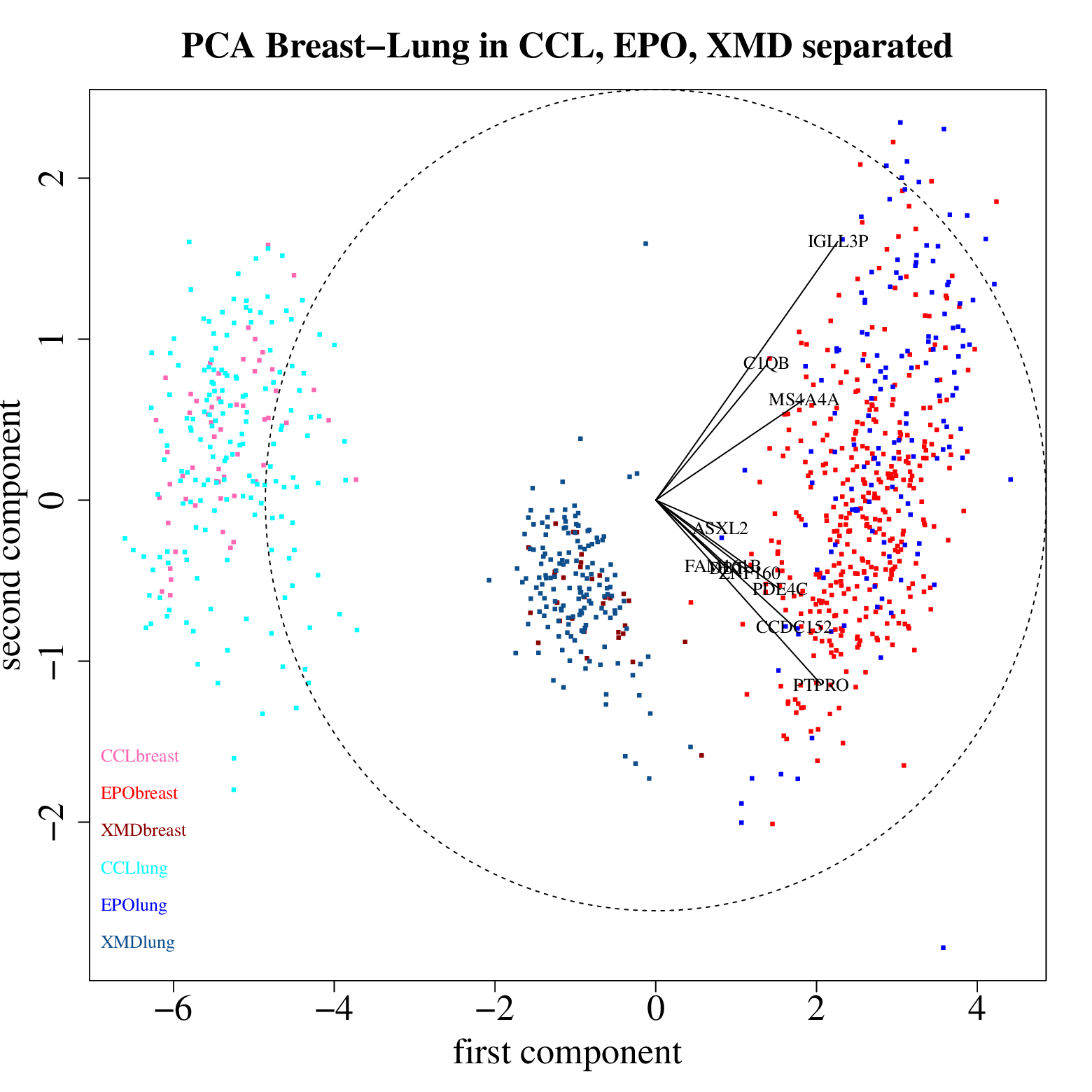}&
\includegraphics[width=8cm]{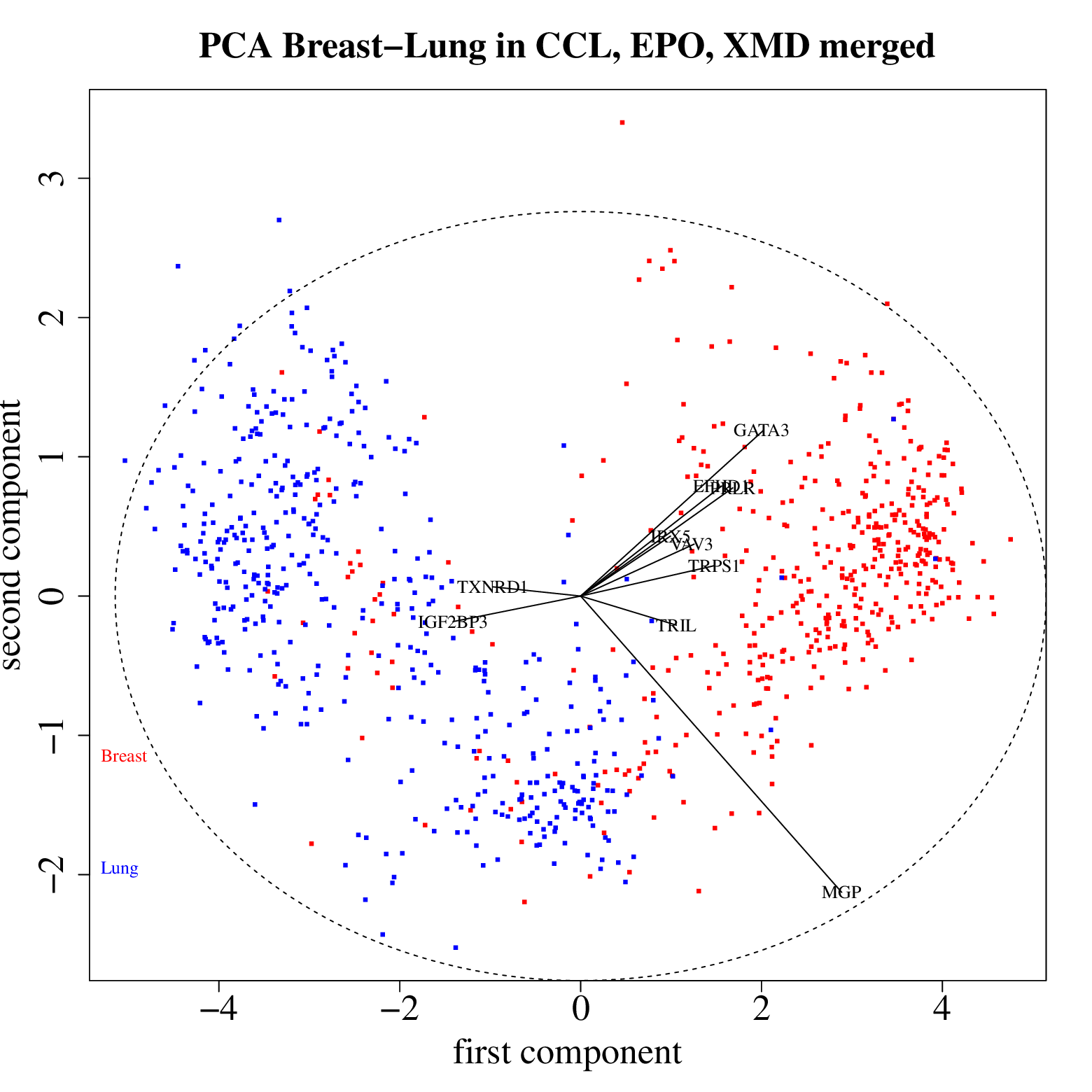}
\end{tabular}
} 
\caption{Principal component analysis of two assays of 10 symbols in
  6 submatrices, extracted from CCL, EPO, and XMD according to
  keywords ``Breast'' and ``Lung''. The six submatrices
  are separated on the right panel, they have been merged on the
  right panel. In each case the 10 most significant features for the
  Kruskal-Wallis test are taken
  as variables. The two sets of 10 symbols are disjoint. Samples are
  represented by red points (Breast) or blue points (Lung).
}
\label{fig:4}
\end{figure}
%-------------------------------------------------------------------
 
Further information was obtained by running 
a one-sided Wilcoxon test to
detect symbols separating both types of tumor. Then the Molecular
Signature database C2
\cite{Subramanian05} was
searched for symbols matching them. Among the 20 genes found most
significantly overexpressed in breast tumors by our test, 11
 were inside genesets of C2
relative to breast cancers, and outside all genesets relative to lung
tumors: 
EFHD1, IRX5, MUCL1, PRLR, PTGER3, RGL2, TRIL, TRPS1, VAV3, WWP1,
ZG16B. Seven of these genes can be found in the G2SBC database
\cite{Mosca10} and for $10$ out of $11$, 
we have found at least one reference relating 
it to breast cancer. 
Conversely, among the most significant genes for lung tumor, 
the following were found in C2 genesets related to lung and not in
those related to breast: ALDH3B1, DARS, PRPSAP2, FAM96B, MBIP, 
LRRC20. The overexpression of ALDH3B1 in lung tumors has been reported in
\cite{Marchitti10}. Santarius et al. \cite{Santariusetal10} gives
lists of genes, the overexpression of which is associated to different
types of human cancers. The genes detected as
significantly overexpressed in Breast by our test, that were also
among class III genes related to breast cancer in Table 1 
of \cite{Santariusetal10},  were  
FGFR1, BAG4, 
MDM2, YWHAB, 
ZNF217. For Lung, they were EGFR, 
MET, YWHAZ, 
MYC, NKX2-1, 
DCUN1D1.
%
% FGFR1 ($P=1.8\times 10^{-27}$), 
% BAG4 ($P=4.6\times 10^{-3}$), 
% MDM2 ($P=1.6\times 10^{-8}$), 
% YWHAB ($P=5.5\times 10^{-4}$), 
% ZNF217 ($P=1.9\times 10^{-18}$). 
%
% For lung, there were EGFR ($P=1.3\times 10^{-3}$), 
% MET ($P=2.8\times 10^{-48}$), 
% YWHAZ ($P=1.1\times 10^{-8}$), 
% MYC ($P=6.5\times 10^{-8}$), 
% NKX2-1 ($P=1.3\times 10^{-45}$), 
% DCUN1D1 ($P=2.8\times 10^{-4}$). 
%
These findings would
require further confirmation over larger datasets. Yet they provide
evidence that meaningful biological information can be extracted by
merging generalist matrices such as CCL, EPO and XMD. 

Our next experiment consisted in merging the two groups of blood RNA
datasets, found to be homogeneous on the correlation analysis (Figure 1): HAV,
HPS, PLE, WBS for healthy individuals, AML, MDS, for leukemias. 
The samples of MIL were separated into MILh (Healthy), and MILl
(leukemias). The left panel of  
Figure 3 shows the first plane of the PCA for
the same 22 features as in Figure 2, the 8 matrices being
represented by different colors. It turns out that the samples
corresponding to MILh are mixed on the representation 
with the other MIL points. Thus they were removed from the matrix
``Healthy'', whereas ``Leukemia'' was made by merging AML, MDS, and
MILl. The Kruskal-Wallis test between Healthy and Leukemia,
detected $16\,977$ significant features out of $17\,691$, among which
$7\,970$ had a null p-value. The right
panel of Figure 5 shows the PCA over 10 of them.
 
%-------------------------------------------------------------------
\begin{figure}[!ht]
\centerline{
\begin{tabular}{cc}
\includegraphics[width=8cm]{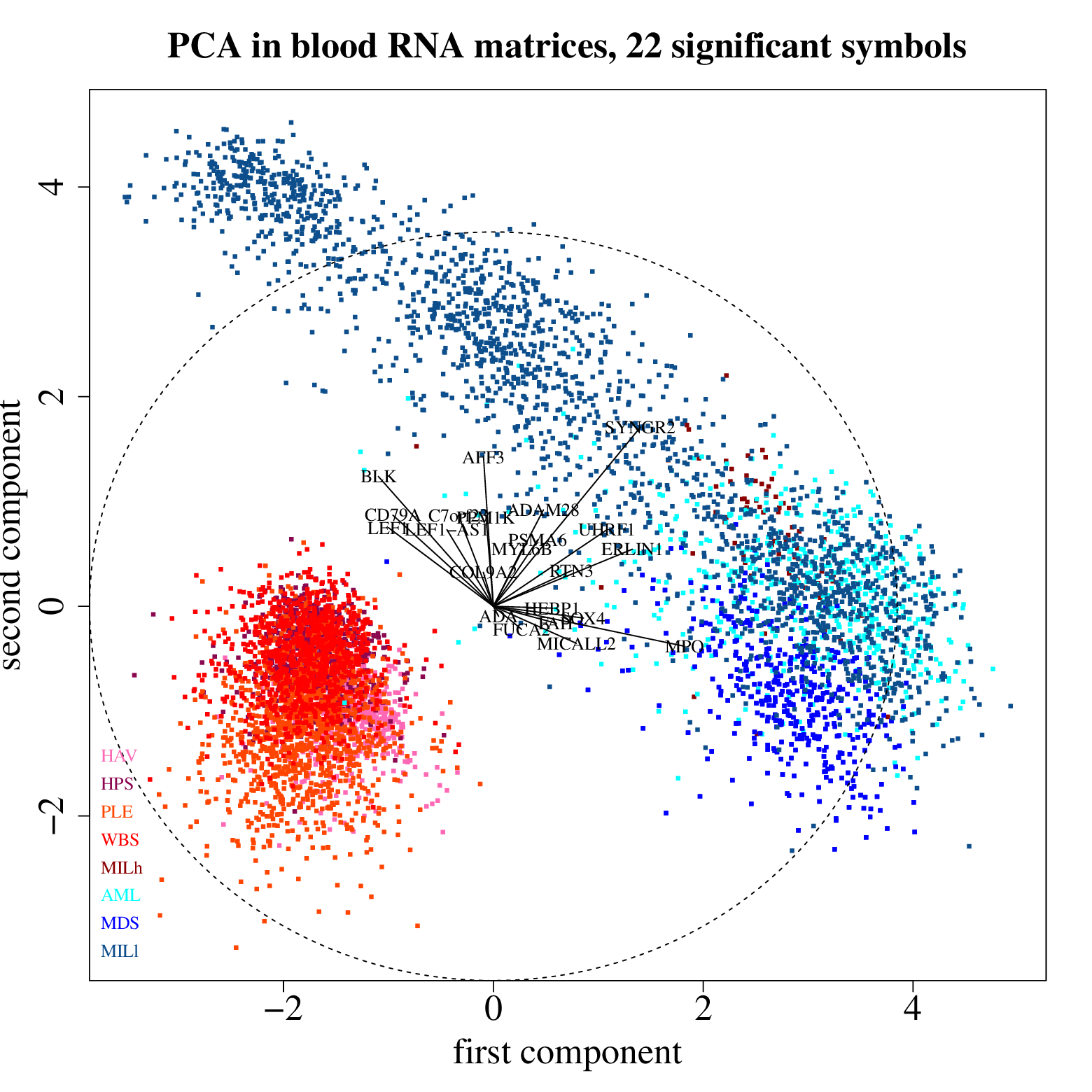}&
\includegraphics[width=8cm]{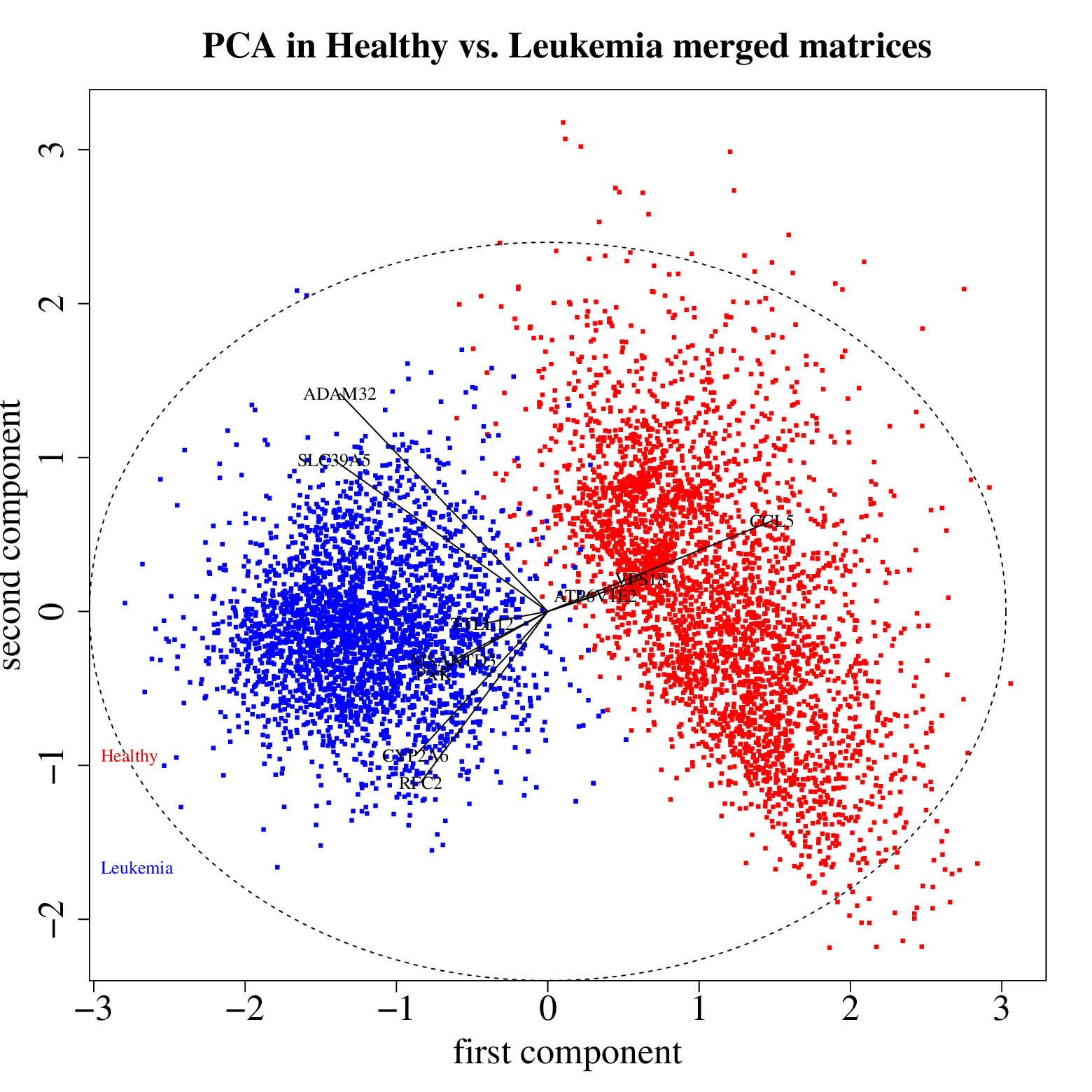}
\end{tabular} 
}
\caption{On the left panel, PCA of 
  HAV, HPS, PLE, WBS, MILh (Healthy = red points), and 
AML, MDS, MILl (Leukemia = blue points), 
for the same 22 features as in Figure 2.
On the right panel, HAV, HPS, ML, WBS, have been grouped into Healthy, 
AML, MDS, MILl into Leukemia. The assay is made of 10
random features among the $7\,970$ having a null Kruskal-Wallis p-value. 
}
\label{fig:5}
\end{figure}
%-------------------------------------------------------------------

 The one-sided Wilcoxon test was run to
detect which symbols were significantly overexpressed in leukemias.
For that test, a set of $4\,191$ symbols had a null p-value. 
A second set of symbol was extracted from C2: those appearing in 
leukemia-related genesets. The C2 set has
$5\,688$ symbols, and the intersection with the first contains
$1\,617$, which is highly significant for Fisher's hypergeometric test
($P=1.36\times 10^{-51}$). The ten symbols found most significant for
leukemia by the Wilcoxon test were RPL34, GABARAP, RPL36A, H2AFV,
CSDE1, DNTTIP2, OPHN1, PABPC3, PNRC1, RPSA. Among those 10, 8 appeared
in the leukemia-related genesets of C2. The symbol H2AFV is found in
six of them. Another noteworthy result concerns the pair of genes
NUP98-TOP1, shown to be related with
leukemia in \cite{Gurevich04}. 
When genes are ranked by decreasing order of significance, 
NUP98 and TOP1 have ranks $187$ and $65$ respectively, which confirms
their link with leukemia.

Another experiment was run on the same matrices, by separating acute
myeloid leukemia samples, from all other
samples. Thus the same calculations that had been run inside MIL
before, were repeated over a larger number of samples. 
The acute myeloid leukemia samples were taken from AML and
MIL ($1\,076$ samples), others were obtained by merging HAV, HPS, PLE, WBS, MDS,
with the non-AML samples of MIL ($6\,010$ samples). The one-sided
Wilcoxon test of comparison was run. For the 7 genes signaled as
overexpressed in AML by \cite{Stirewaltetal08}, the results were much
more significant as before: the least significant p-value was that of BIK:
$8.4\times 10^{-35}$, whereas 
FUT4 and HOMER3 had p-values below machine
precision. Contrarily to the study that had been conducted inside MIL,
a clear confirmation was also obtained for 
the genes reported by \cite{Stirewaltetal08} to be underexpressed
in case of AML. Five of them were in the common features of our
matrices, four had p-values smaller than $10^{-100}$ for
underexpression in AML. In particular, PELO and PLXNC1 who had not
been found significantly underexpressed in the first experiment, now had
p-values $3.5\times 10^{-238}$ and  $4.4\times10^{-168}$ in the test on merged
matrices.

%-------------------------------------------------------------------
 
%
\section*{Discussion}
A new set of R functions has been
developed. Like other packages  \cite{Davis07,Taminau11}, 
it performs the usual formatting operations. It also offers new
functionalities for sorting lists of datasets according to information
keywords. Various robust statistics techniques are encoded.
The script and a user manual have been made available online.
Using these R functions, a large scale study of $20$ GEO datasets,
totalling $17\,745$ 
samples, has been conducted. 

Our first
conclusion is that Haibe-Kains et al. \cite{Haibe-Kains13} were right
in observing that 
inconsistencies between datasets make it dangerous to merge them
without precautions. The risk is to declare as biologically
significant, observations which are actually statistical artifacts.
The first precaution is to transform the data into
distribution-free values, i.e. to use robust statistics. 
This implies replacing the data of each sample
by their empirical distribution function, or some other
distribution-free score \cite{GibbonsChakraborti03,Heritieretal09}. 
Even after data have been homogenized, important
discrepancies remain. 
For this reason, checking comparability between studies before merging
them is imperative. 
One possible measure of similarity
(among others, see for instance \cite{Parmigiani04}) for two datasets is 
the correlation between medians, which has been used here. 
Two sets of samples corresponding to different
conditions inside one given homogeneous 
dataset usually have correlations of
medians above $0.8$ (see Table 3). Arguably, it can be considered
that two different datasets can safely be merged, if all
paired-correlations between medians
are above $0.8$. This is not always the case, even
between datasets coming from the same tissues, obtained under the same
platform (see Table 2). Further ways of investigating possible
discrepancies involve multivariate statistics. 
Graphical methods include Factor Analysis, Principal Component
Analysis, Discriminant Analysis \cite{Hardleetal12}. Inference can be done
using the robust equivalents of usual normal-sample methods,
i.e. Wilcoxon test instead of Student's t-test, Kruskal-Wallis instead
of one-way anova, etc. When repeatedly applying such a test to a 
set of symbols, a False Detection Rate (FDR) correction
must be applied to the p-values. We have chosen the Benjamini-Yekutieli method
\cite{Benjamini_Yekutieli01}. 
Our observation was that, even after FDR correction, the
tests usually detect a sizeable proportion of
all symbols as significant for discrimination, either between several
different datasets, or between different types of samples within the
same dataset. We believe that relevant biological information
can be obtained from applying a discriminating test, 
then ranking features according to their degree of significance,
i.e. ordering the values obtained over each feature by the test
statistic. In the cases considered here (breast tumors against lung
tumors, healthy blood samples against leukemias, acute myeloid
leukemia against other blood RNA samples), it was observed that
among the most significant symbols, a large proportion of them
were already known as being related to the corresponding cancers. 
This can be viewed as evidence that meaningful biological information
can be extracted by merging different datasets.
We believe that important new findings could be obtained
by the same method, being aware that a statistical listing of
significant symbols does
not necessarily imply that all listed symbols correspond to
true biological information. Such a list must necessarily be expert-curated
for biochemical validation.

%%%%%%%%%%%%%%%%%%%%%%%%%%%
\section*{Acknowledgements}
  \ifthenelse{\boolean{publ}}{\small}{}
  BY, KC, and JJF acknowledge financial support from 
Laboratoire d'Excellence TOUCAN (Toulouse
  Cancer). 
%%  Bmc_article.bst  will be used to                       %%
%%  create a .BBL file for submission, which includes      %%
%%  XML structured for BMC.                                %%
%%  After submission of the .TEX file,                     %%
%%  you will be prompted to submit your .BBL file.         %%

%%%%%%%%%%%%%%%%%%%%%%%%%%%%%%%%%%%
%%                               %%
%% Additional Files              %%
%%                               %%
%%%%%%%%%%%%%%%%%%%%%%%%%%%%%%%%%%%

\section*{Additional Files}
Additional material has been provided as a compressed directory
available online:\\
\verb+http://ljk.imag.fr/membres/Bernard.Ycart/publis/sagd.tgz+\\
It contains:
\begin{enumerate}
\item a R script file \texttt{sagd.r}: the R functions implementing the
  method described here,
\item a pdf file \texttt{sagd\_manual.pdf}: user manual for the R
  functions.
\end{enumerate}

%  \subsection*{Additional file 1 --- Sample additional file title}
%    Additional file descriptions text (including details of how to
%    view the file, if it is in a non-standard format or the file extension).  This might
%    refer to a multi-page table or a figure.
%
%  \subsection*{Additional file 2 --- Sample additional file title}
%    Additional file descriptions text.

%\newpage
{\ifthenelse{\boolean{publ}}{\footnotesize}{\small}
 \bibliographystyle{bmc_article}  % Style BST file
\bibliography{sagd.bib} }
%\bibliography{/home/ycart/recherche/Fournie/JJ.bib} } 
% Bibliography file (usually '*.bib' ) 

%%%%%%%%%%%

\ifthenelse{\boolean{publ}}{\end{multicols}}{}

%%%%%%%%%%%%%%%%%%%%%%%%%%%%%%%%%%%
%%                               %%
%% Figures                       %%
%%                               %%
%% NB: this is for captions and  %%
%% Titles. All graphics must be  %%
%% submitted separately and NOT  %%
%% included in the Tex document  %%
%%                               %%
%%%%%%%%%%%%%%%%%%%%%%%%%%%%%%%%%%%

%%
%% Do not use \listoffigures as most will included as separate files
%-------------------------------------------------------------------
%-------------------------------------------------------------------
%
%-------------------------------------------------------------------

%%%%%%%%%%%%%%%%%%%%%%%%%%%%%%%%%%%
%%                               %%
%% Tables                        %%
%%                               %%
%%%%%%%%%%%%%%%%%%%%%%%%%%%%%%%%%%%

%% Use of \listoftables is discouraged.
%%

\end{bmcformat}
\end{document}